\documentclass[aps,pra,twocolumn,groupedaddress]{revtex4-1}

\usepackage{amssymb,amsfonts,amsmath}

\usepackage{graphicx}
\usepackage{graphics}
\usepackage{amssymb}
\usepackage{epstopdf}

\renewcommand{\r}{{\bf r}}

\begin{document}


\title{Quantum Collapse and the Second Law of Thermodynamics}


\author{Sahand Hormoz}
\email[]{hormoz@kitp.ucsb.edu}
\affiliation{Kavli Institute for Theoretical Physics, Kohn Hall, University of California, Santa Barbara, CA 93106, USA}


\renewcommand{\today}{August 11, 2012}
\date{August 11, 2012}

\begin{abstract}
A heat engine undergoes a cyclic operation while in equilibrium with the net result of conversion of heat into work. Quantum effects such as superposition of states can improve an engine's efficiency by breaking detailed balance, but this improvement comes at a cost due to excess entropy generated from collapse of superpositions on measurement. We quantify these competing facets for a quantum ratchet comprised of an ensemble of pairs of interacting two-level atoms. We suggest that the measurement postulate of quantum mechanics is intricately connected to the second law of thermodynamics. More precisely, if quantum collapse is not inherently random, then the second law of thermodynamics can be violated. Our results challenge the conventional approach of simply quantifying quantum correlations as a thermodynamic work deficit.
\end{abstract}

\pacs{}
\keywords{Quantum Collapse | Quantum Measurement | Second Law of Thermodynamics | Maxwell Demon}

\maketitle

\section{Introduction}
Kelvin's statement of the second law of thermodynamics claims that ``no process is possible in which the sole result is the absorption of heat from a reservoir and its complete conversion into work." \cite{Kelvin} First written in 1849, more than a half-century before the discovery of quantum mechanics, it is generally regarded as one of the immutable laws of nature. Here, we explore the implication of this statement on the laws of quantum mechanics, in particular the measurement postulate.

According to quantum theory, on measurement a state collapses probabilistically into an eigenstate of the measured observable \cite{Bohr,Born}. The \emph{objectively} random outcomes from measurements of a superposition of eigenstates have been proposed as an unbiased physical source for generating random numbers \cite{Zeilinger00}. Thus far, experimental tests have found no deviation from randomness for sequences generated using quantum collapse \cite{ExpTestRand1,ExpTestRand2,ExpTestRand3}. However, despite some limited theoretical constraints on computability of quantum measurement outcomes \cite{Calude08}, the inherently probabilistic nature of quantum collapse remains a postulate; for a derivation of collapse probabilities without appealing to Born rule see \cite{Zurek09a,Zurek09b}. We suggest a possible connection between the inherent indeterminacy in quantum collapse and the second law of thermodynamics by exploring quantum protocols for converting heat to work.

A heat engine is a system that is cyclically modified while in thermal equilibrium with the net result of conversion of heat into mechanical work. To extend this notion into the quantum regime, quantum analogues of various isothermal and adiabatic steps for running the engine have been considered \cite{ZurekSzilard,Kieu04,Nori07,Ueda11}; other quantum effects such as the discrete nature of energy states \cite{Kieu04} and indistinguishability of particles \cite{Ueda11} can also modify the operations of a quantum engine. However, more exotic quantum properties such as quantum coherence, at first seem impossible to achieve in a system in contact with a thermal reservoir.  Thermal equilibrium implies that the engine is described by a canonical ensemble of orthogonal energy eigenstates --diagonal density matrix. Nonetheless, if some degrees of freedom of the system are neglected (traced over), the remaining subsystem is not in general described by a canonical ensemble. To access exotic quantum features, the observer operating the engine must focus on a subsystem. A bipartite system shared between two parties is the simplest quantum engine of interest.

In \cite{Oppenheim02} and \cite{Zurek03}, it is shown that less work can be extracted from a heat bath when a bipartite system is shared between two parties than when one party has global possession of the system. Thermodynamic measures such as the amount of extractable work are used to quantify available resources for quantum information processing, such as quantum correlations. However, despite their utility in providing a quantitative metric for non-local resources, these approaches do not estimate the maximum locally extractable work correctly, by missing a crucial ingredient: the same non-orthogonal states that introduce added inefficiencies in the operations of a local observer can also break detailed balance. The subsystems are effectively out of equilibrium, and measures deduced from equilibrium thermodynamics are not strictly valid.

Here, we quantify the thermodynamic benefits and costs of locally manipulating a simple quantum heat engine comprised of an ensemble of interacting pairs of atoms. We estimate the locally extractable work by accounting for the broken detailed balance associated with the collective measurement of a subsystem (one atom of each pair). With detailed balance broken, the engine can rectify thermal fluctuations to extract work, much like the ``ratchet and pawl'' engine used by Feynman in his {\it Lectures} \cite{Feynman}. In the proposed quantum ratchet, irreversibility of quantum collapse (or decoherence) permits extraction of net motion from thermal fluctuations of a single heat bath. We show that the work cost due to the excess entropy generated from the collapse of non-orthogonal states, compensates for the gain obtained from ratcheting thermal fluctuations. Kelvin's statement is never violated. We use this simple construct to demonstrate that if a sequence of collapsed measurement outcomes is nonrandom, or algorithmically compressible, then the second law of thermodynamics can be violated. 

Since we are interested in measurements and the role of an observer in extracting thermodynamic work, we adopt the setup of Maxwell's demon \cite{Maxwell}. This demon can measure the state of a system, for example the position of a single gas molecule in a chamber, and extract work from its knowledge, for instance by positioning a piston appropriately and carrying out an isothermal expansion \cite{Szilard}. As observed by Landauer and Bennett \cite{LBa,LBb}, the demon does not violate Kelvin's statement, since in a cyclic operation the same amount of work is required to erase the demon's memory as is extracted. Maxwell's demon establishes a connection between information and thermodynamic entropy. The flow of information is analogous to the flow of heat in and out of a Carnot engine from thermal reservoirs. If the reservoirs have the same temperature, the inflow and outflow of heat are the same, and the net work extracted zero. Implications of a demon measuring quantum states were considered early on by von Neumann \cite{vonNeumann55}, who showed that the ability to distinguish non-orthogonal states is equivalent to a violation of the second law.

At first sight, quantum measurements seem to introduce further inefficiencies in the operations of the demon. Take the example of a single spin in an external magnetic field first discussed in \cite{Lloyd97}. The spin is initially in state $\mid \rightarrow \rangle = 1/\sqrt{2}(\mid \uparrow \rangle + \mid \downarrow \rangle)$, where $\mid \downarrow \rangle$ ($\mid \uparrow \rangle$) points in the same (opposite) direction as the external field and has energy $-\mu B$ ($\mu B$). One way to extract work from such system is to rotate the spin to the $\mid \downarrow \rangle$ by applying a $\pi/2$ pulse, extracting work $\mu B$. Alternatively, one can measure the spin along the magnetic field direction, if the outcome is $\mid \uparrow \rangle$ the spin can be rotated to $\mid \downarrow \rangle$ state with a $\pi$ pulse to extract work $2\mu B$. Since this outcome occurs half the time, on average the work extracted is $\mu B$, same as before. However, the process of measurement has generated a `waste' bit of information, which requires work $k_B T \ln 2$ to erase, making the overall process less efficient. Naively, one might expect that quantum effects in general can only decrease efficiency by generating excess entropy from measurements of non-orthogonal states.

Scully et al. in a pioneering paper \cite{Scully03} showed how work can be extracted from a single \emph{quantum} heat bath. The key idea is to use quantum coherence to break detailed balance. Using a three-level atomic system in a photon bath, they showed that a properly tuned super-position of the almost-degenerate ground states can result in destructive interference between the absorption paths. With a reduced absorption probability, the atoms can effectively act as a higher temperature reservoir, permitting extraction of work. Of course, the cost of generating the initial coherence precludes the possibility of violating the second law.

Here, we combine the above seemingly contrasting views into a consistent picture using the language of Maxwell's demon. Before introducing the specifics of the quantum ratchet, we consider a general cycle for a quantum heat engine. 

\section{Generic engine cycle}
Consider the following cyclic operation for an engine with access to a single heat bath; for similar cycles see \cite{ZurekSzilard,Nori07,Kieu04,Ueda08}\\
\emph{Step 1--} The system with Hamiltonian $H_1$ is put in contact with a thermal reservoir at temperature $T$. At equilibrium system is descried by density matrix $\rho_1 = \frac{1}{Z_1} e^{- \beta H_1}$, where $\beta = (k_B T)^{-1}$ and $Z_i = Tr\{e^{-H_i/k_B T}\}$ is the partition function.\\
\emph{Step 2--} The system is isolated from the thermal reservoir. The Hamiltonian is \emph{adiabatically} changed to $H_2$, leaving the occupation probability of each energy level unchanged. The density matrix of the system after the transition is $\bar{\rho}_1 = \sum_i p_i |e_i \rangle \langle e_i |$, where $\{ p_i \}$ are the eigenvalues of $\rho_1$ and $\{ | e_i \rangle \}$ the energy eigenstates of $H_2$ \cite{Nori07}. The amount of work performed \emph{by} the system is $W_1 = Tr\{ \rho_1 H_1\} - Tr\{ \bar{\rho}_1 H_2\}$ \cite{Kieu04}.\\
\emph{Step 3--} The demon extracts work from the system by first measuring it and gaining information $S(\bar{\rho}_1) = S(\rho_1) =   -Tr \{ \rho_1 \ln \rho_1 \}$, where $S$ is the von Neumann entropy. The system is then put in contact with the thermal bath and work extracted from a quasi-static isothermal expansion in phase space from the known state (measurement outcome) to one with maximum entropy \cite{LBa,LBb,Zurek03,Ueda08}; the final density matrix is $\rho_2 = \frac{1}{Z_2} e^{- \beta H_2}$. Using the first law of thermodynamics, work done by the system is given by,
\begin{equation}
W_3 = \Delta U_3 - \beta^{-1} Tr \{\rho_2 \ln \rho_2 \} + \beta^{-1} Tr \{ \rho_1 \ln \rho_1 \},
\end{equation}
where $\Delta U_3 = Tr \{ \bar{\rho}_1 H_2 \} - Tr \{ \rho_2 H_2 \}$ is the change in internal energy of the system. The second term on the right hand side the is heat flow from the isothermal expansion, and the last term, the work cost of erasing the information gained from the demon's measurement.\\
\emph{Step 4--} Finally, to complete the cycle the Hamiltonian is adiabatically changed back to $H_1$ with the work extracted given by $W_4 = Tr\{ \rho_2 H_2\} - Tr\{ \bar{\rho}_2 H_1\}$, where $\bar{\rho}_2$ has the same occupation probabilities as $\rho_2$ but with eigenstates corresponding to the energy levels of $H_1$. 

The net work by the system in the full cycle is,
\begin{equation}
W_{net} = Tr \{ (\rho_1 -\bar{\rho}_2) H_1\} + \beta^{-1} \big( S(\rho_2) - S(\rho_1) \big). \label{WnetGlobal}
\end{equation}
Substituting $H_1 = -\beta^{-1}\ln(Z_1 \rho_1)$ in above expression gives $W_{net} = \beta^{-1} (Tr \{\bar{\rho}_2 \ln \rho_1 \} - Tr \{ \rho_2 \ln \rho_2 \})$. Since $S(\rho_2) \leq -Tr\{\bar{\rho}_2 ln \rho_1 \}$ \cite{Vitelli01}, for any choice of $H_{1,2}$, $W_{net} \leq 0$, which implies that the cycle can only convert work to heat, in agreement with Kelvin's statement.


The demon, however, can access non-equilibrium ensembles by observing subsystems. The broken detailed balance in the subsystems can be exploited to enhance efficiency.  However, observing subsystems can generate excess entropy and increase the work cost of erasure \cite{Zurek03}, which we demonstrate first. Assume that there are two subsystems, A and B, with reduced density matrices $\rho^{A(B)} = Tr^{B(A)} \{ \rho \}$. Step 3 above needs to be modified to,\\
{\emph Step 3'--} The demon measures subsystem A, gaining information $S(\rho^A_1) = -Tr \{\rho^A_1 \ln \rho^A_1 \}$. The system is then put in contact with the thermal bath and work extracted as before. The demon does the same with subsystem B. Information gained from the second measurement, however, is not independent from the first one, and given by,
\begin{equation}
 S(\rho^B_1|\Pi_A) = \Sigma_i p_{\pi_A = a_i} S(\rho^{B|\pi_{a_i}}_1),
\end{equation}
where $\Pi_A$ is a complete projective measurement on A; $p_{\pi_A = a_i} = Tr\{\rho^A_1 \Pi_{A=a_i}\}$ is the probability of outcome $a_i$ from measuring subsystem A; and $\rho^{B|\pi_{a_i}}_1 = Tr^A\{\rho_1  \Pi_{A=a_i} \}/p_{\pi_A = a_i}$ is the state of B after the measurement on A.

The net work of one cycle is,
\begin{equation}
W'_{net} = Tr \{ (\rho_1 -\bar{\rho}_2) H_1\} + \beta^{-1} \big( S(\rho_2) - S(\rho^A_1) -  S(\rho^B_1|\Pi_A) \big). \label{WnetPart}
\end{equation}

Local measurement can generate additional entropy, $S(\rho^A_1) + S(\rho^B_1|\Pi_A) > S(\rho_1)$, if there are quantum correlations, for example entanglement, present between the two subsystems \cite{Zurek01a,Zurek01b,Vedral01}. Quantum correlations exist even if the individual states making up $\rho$ are separable, $\rho = p_{i,j} \vert A_i \rangle \langle A_i \vert \otimes \vert B_j \rangle  \langle B_j \vert$, as long as states $\vert A_i \rangle$ ($\vert B_j \rangle$) form a non-orthogonal set for subsystem A (B) \cite{Vedral10b}. Quantum correlations are encoded in non-orthogonal states of the subsystems or entangled states, which collapse when measured locally. The excess entropy generated from the measurement-induced collapse of these states is given by,
\begin{equation}
\delta = S(\rho^A_1) + S(\rho^B_1|\Pi_A) - S(\rho_1). \label{discord}
\end{equation}
When minimized over all measurement basis $\Pi^A$, this quantity is known as quantum discord \cite{Zurek01a,Zurek01b,Vedral10}. $\delta$ is always greater than or equal to zero.

The net work extracted from local measurements Eq. \eqref{WnetPart} can be naively related to that from global measurements Eq.\eqref{WnetGlobal}, $W'_{net} = W_{net} - \beta^{-1}\delta$; as argued in \cite{Oppenheim02,Zurek03}. However, Eq.\eqref{WnetPart} has neglected a crucial ingredient: the same quantum correlations that create discord can also break detailed balance. Measuring subsystems, besides generating excess erasure cost, can also enhance engine efficiency. To correctly calculate the maximum extractable work, we need to exploit the broken detailed balance in the engine.

\section{Quantum ratchet}
Periodic potentials with broken parity symmetry (e.g. a sawtooth pattern), might seem capable of extracting net directed motion (and thereby work) by rectifying thermal kicks. Smoluchowski and later Feynman, using a ``ratchet and pawl'' construct, showed that detailed balance precludes this possibility \cite{Smoluchowski,Feynman}. Out of equilibrium, however, correlated fluctuations \cite{Magnasco}, non-thermal kicks, or time-varying potentials \cite{Astumian}, can give rise to directed motion.

We use the inherent non-equilibrium nature of quantum collapse to implement a quantum ratchet. But first, consider a simple equilibrium case: an ensemble of two-level atoms in contact with a thermal reservoir at temperate $T$ (Fig.1a); a fraction of the atoms are thermally excited. Assume that a pulse of light exists that only excites the atoms in the ground state, leaving the excited atoms unaltered. After applying such a pulse, which requires work, all the atoms will be in the excited state (Fig.1b). The demon then extracts work by rotating all the atoms back to the ground state. Since the entropy of the system is zero after the pulse, there is no erasure work. The net work extracted is equal to the energy of the thermally excited atoms, as the work extracted from the atoms initially in the ground state is equal to the work spent applying the pulse. This protocol `ratchets' work from thermal excitations of the system. Of course, such a pulse does not exist, as it violates detailed balance: absorption and emission are no longer on the same footing. A $\pi$ pulse, for instance, will also result in decay of all the excited atoms in addition to exciting all the ground state atoms. More generally, unitarity in quantum mechanics requires that any coupling be hermitian, implying that any transfer from the ground state to the excited state must occur equally in the reverse direction.

\begin{figure}
\begin{center}
      \centerline{\includegraphics{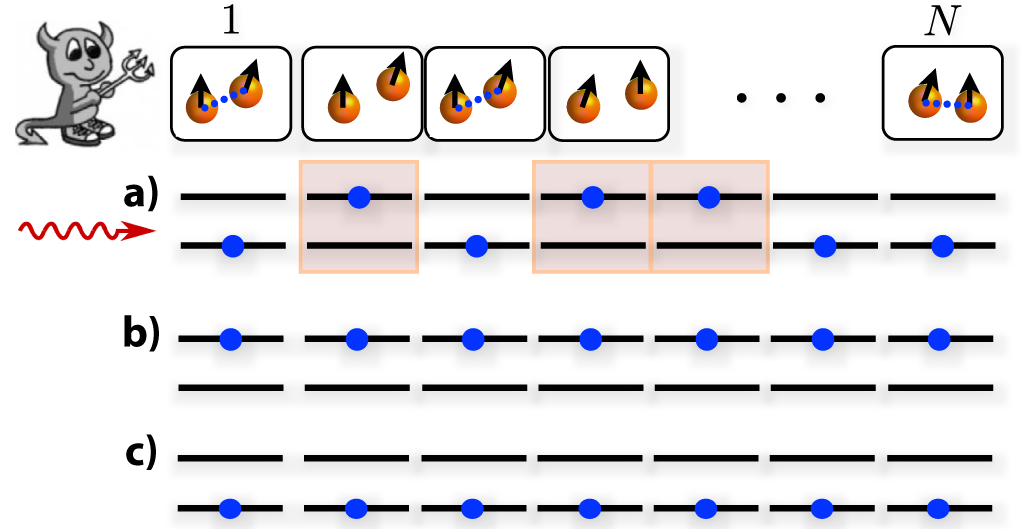}}
       \caption{Quantum Ratchet. a) An ensemble of two level systems in equilibrium at temperature T. Application of the pulse excites the atoms that were initially in the ground state, resulting in (b). Atoms in the excited state (highlighted) are not altered. Work is extracted by the demon rotating all the atoms to the ground state (c). Such a pulse violates the principle of detailed balance. However, a physical realization is possible using a collective quantum non-demolition measurement of the entangled pairs of atoms.} \label{Fig1}
\end{center}
\end{figure}

Nonetheless, it is possible to implement the above ratchet using entangled pairs of atoms as effective two-level systems. Our engine will be an ensemble of $N \gg 1$ pairs of interacting atoms (A and B) each in a box of size $L$. The internal state of the atoms is represented as a spin-$1/2$ system with energy gap $\omega$. The two atoms interact (XY-interaction) with a coupling constant that is a function of their separation distance.

The Hamiltonian for each subsystem (pair of atoms) is given by,
\begin{equation}
H = \omega S^z_A + \omega S^z_B + \lambda(\r)(S^+_AS^-_B + S^-_AS^+_B), \label{atomH}
\end{equation}
where $S^z_j = \sigma^z_j/2$ and $S^{\pm}_j = S^x_j \pm i S^y_j$, with $\sigma^{x,y,z}_j$ the Pauli matrix acting on atom $j=A,B$.

The coupling is only turned on when the separation distance of the atoms is below a certain cutoff.
\begin{equation}
\lambda(\r) = \begin{cases} \lambda, & \mbox{if } |\r| \leq r_0 \\ 0, & \mbox{if } |\r|>r_0 \end{cases}
\end{equation}

The engine is operated as follows (for a detailed analysis see Supplementary Information):\\
\emph{Step 1--} Initially, the box size $L$ is smaller than the cutoff $r_0$. Each pair is equilibrated via exchange of photons with an incoherent thermal radiation field. Above Hamiltonian has four energy eigenstates; two separable states: $|gg\rangle$ and $|ee\rangle$,where $|e\rangle$ and $|g\rangle$ denote the single-atom excited and ground states ($S^z=\pm1/2$); two entangled energy eigenstates: $|\pm \rangle = 1/\sqrt{2}(|eg\rangle \pm |ge\rangle)$, which have an energy separation of $2\lambda$. The equilibrium occurrence probabilities of the entangled states are denoted as $p_\pm$.

\emph{Step 2--} The radiation field that thermalized the internal state of each atom-pair is turned off, ensuring that the probability of occurrence of each internal state is kept constant. The translational degrees of freedom of the atoms, however, can equilibrate via collisions with the walls of the box. The volume of the boxes are doubled. Atom-pairs in the separable states have no energetic preference for their separation distance. The energy of the entangled state $|-\rangle$, however, increases by $\lambda$ if the separation distance of the atoms exceeds the cutoff (uncoupling the pair); thermal fluctuations can provide this energy. Conversely, state $|+\rangle$ has a lower energy in the uncoupled state. The equilibrium probability of finding internal states $|\pm\rangle$ in the uncoupled state is $p^\pm_o = 1/(1+e^{\mp \beta \lambda})$ (Fig.2).

\emph{Step 3--} A quantum non-demolition measurement is performed on the collective spin of all the A-atoms, $\tilde{S}^z_A = 1/\sqrt{N/2} \sum_{i=1}^N S_{A,i}^z$, where the $i$ summation runs over all the boxes. The measurement outcome does not reveal which A-atoms are excited, but only their total number \cite{Lukin}. The information obtained is logarithmic in the system size $N$ and has a negligible erasure cost. The measurement collapses all the entangled states, since a superposition of an A-atom in the excited and ground states can not correspond to a known total number of excited A-atoms. (Non-demolition means that the measurement Hamiltonian commutes with the single-atom Hamiltonian ($S_A^z$), leaving the A-atom states unaltered; see Supplementary Information for an implementation using Faraday-rotation). The average energy of a coupled atom-pair increases to their uncoupled state after the collapse. This is manifested in the higher effective temperature of the individual atoms (Fig.2 inset); work is required for the collective measurement. However, as in the simple example above, this work will be retrieved in the final step. For the pairs thermally excited to the uncoupled state, the collapse requires no work. Restoring these pairs to the initial state will effectively convert thermal excitations in the translational motion of the atoms to useful work; the cycle operates like a quantum ratchet. Unlike the simple $\pi$ pulse of the above example, the collective measurement inherently breaks detailed balance. Decoherence/collapse of the entangled states is an irreversible process. The measurement increases the energy of the coupled pairs to that of uncoupled pairs, but not the reverse.

\emph{Step 4--} The volume of the boxes are restored to their initial value. The demon measures the internal state of each pair of atoms, reducing their entropy to zero. An isothermal reversible expansion is performed to the initial state. As discussed above, the probabilistic collapse of the entangled states, by the local (one atom of each pair) collective measurement, results in an increase in entropy by $\delta$. Therefore, the erasure cost will exceed the work extracted during the isothermal expansion by $\beta^{-1}\delta$. The ratchet provides useful work by effectively lowering the work cost of the collective measurement in the previous step. Thermally uncoupled $|-\rangle$ states save $\lambda$ in measurement work, whereas, uncoupled $|+\rangle$ states require additional work $\lambda$. On average, the net work extracted in the cycle from each atom-pair is
\begin{equation}
W''_{net} = \lambda(p_- p^-_o - p_+ p^+_o) - \beta^{-1}\delta, \label{WnetFinal}
\end{equation}
where the first term on the right hand side is the work obtained from ratcheting thermal fluctuations in the translational motion of the atoms, and the second term, the erasure cost of the excess entropy generated from the random collapse of the entangled states.

Measurement-induced collapse of quantum states permits ratcheting of work from thermal fluctuations by breaking detailed balance. Above, this has increased the net extracted work compared to the protocol in Eq. \ref{WnetPart}, where $W'_{net} = -\beta^{-1}\delta$ with local measurements but without a ratchet.

The erasure cost of $\delta$ precludes the possibility of violating the second law of thermodynamics. In the above cycle, $W''_{net} \leq 0$, for all values of coupling $\lambda$ and temperature $T$ (Fig.2).  Unlike Eq. \ref{WnetPart}, local measurements of subsystems do not just introduce added inefficiencies through $\delta$, but also enhance efficiency by breaking detailed balance. Non-random quantum collapse will have thermodynamical implications that we explore below.

\begin{figure}
\begin{center}
      \centerline{\includegraphics{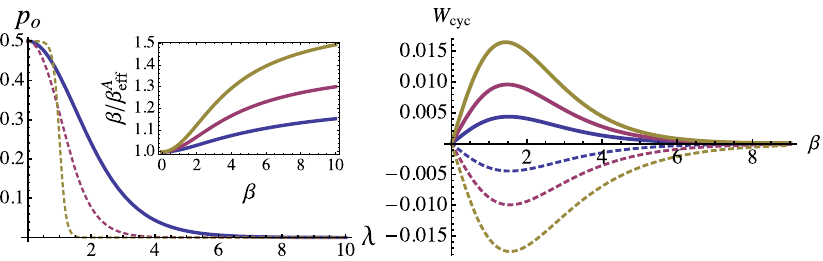}}
       \caption{Net work extracted. Left: probability of finding the pair of atoms in the uncoupled state as a function of coupling parameter $\lambda$, for, from right to left, $\beta=1$,$2$, and $10$; $\omega=1$. The inset shows the ratio of the effective temperature of each atom to the actual temperature as a function of inverse temperature, for, from bottom to top, $\lambda=0.2$, $0.3$ and $0.4$. Right: Net work extracted from the cycle for the same color-coded values of $\lambda$ as the inset. The three curves on the top correspond to $\delta=0$ and imply a violation of Kelvin's statement.} \label{Fig2}
\end{center}
\end{figure}

\section{Non-random collapse}
Each cycle of our engine produces a classical measurement outcome stored by the demon. Quantum theory tells us the occurrence probability of a particular outcome $p_i$ \cite{Born,Zurek09a,Zurek09b}. Each outcome is assumed to be uncorrelated from the preceding ones; the indeterminacy of each measurement is an inherent property of the system. The demon stores a sequence of its measurement outcomes, $s$. In the limit $N \to \infty$ outcomes, this sequence can be reversibly compressed into a sequence of size $N\mathcal{H}(\{ p_i \})$, where $\mathcal{H}$ is the Shannon entropy of outcome probabilities $p_i$. The compressed sequence must be irreversibly erased to complete the cyclic requirement of the engine.

Of course, the demon does not need to know about quantum theory. It simply compiles a large sequence of measurement outcomes and then searches for a method to maximally compress this sequence to minimize the work for erasure. In general, the compressed sequence $s^*$ can be thought of as a program that when executed on a universal computer, yields $s$. The size of this program $|s^*|$ is called the Kolmogorov complexity of $s$, $K(s)$ \cite{Kolmogorova,Kolmogorovb,Kolmogorovc}, and corresponds to the minimum number of bits that need to be irreversibly erased, setting the minimum erasure work \cite{Zurek89}.

The average Kolmogorov complexity of all possible sequences of outcomes is well-approximated by the entropy of the sequences, $\langle K(s) \rangle_s \approx N\mathcal{H}(\{ p_i \} )$; almost all sequences are random strings and algorithmically incompressible \cite{Kolmogorova,Kolmogorovb,Kolmogorovc}. However, an infinitesimal fraction of the sequences have concise descriptions that are asymptotically negligible compared to the size of the sequence, $\lim_{N \to \infty} K(s)/N \to 0$. Such strings, like the binary digits of $\pi$, might appear random but have concise algorithmic descriptions. $K(s)$ is not a computable function \cite{Kolmogorova,Kolmogorovb,Kolmogorovc}; it is impossible to distinguish between random strings and those that just `look' random. An observer making a finite number of measurements and performing a finite number of statistical tests can never be sure if the outcomes are truly random; for experimental searches for non-randomness in quantum collapses see \cite{ExpTestRand1,ExpTestRand2,ExpTestRand3}. Below, we show that an algorithmically compressible or non-random sequence of quantum measurement outcomes will result in a violation of the second law of thermodynamics.

Assume that quantum measurement outcomes are deterministic, that is the observer can algorithmically compute the expected outcome of a collapse based on its observations. Consider a bipartite ensemble with quantum correlations described by density matrix $\Sigma_i p_i \vert A_i \rangle \langle A_i \vert \otimes \vert B_i \rangle  \langle B_i \vert$. The joint-states $\vert A_i \rangle \otimes \vert B_i \rangle$ form an orthogonal set, however, for subsystem A(B), $\{ \vert A_i \rangle \}$($\{ \vert B_i \rangle \}$) are not necessarily orthogonal. The entropy of this system is $\mathcal{H}(\{ p_i \} )$. Local measurements of subsystem $A$ are assumed to be deterministic, such that the subset of non-orthogonal states $\vert A_{i \in k} \rangle$ that are expected to collapse with outcome $k$ are known a priori. The probability of measuring outcome $k$ is $p_k = \Sigma_{i \in k} p_i$. A subsequent measurement of $B$ fully identifies the initial state. The entropy of the outcomes from local measurement of $A$ followed by $B$ is given by $\mathcal{H}(\{ p_k \} ) + \Sigma_k p_k \mathcal{H} ( \{ p_{i \in k} \} ) = \mathcal{H} (\{ p_i \})$. The equality follows from additivity of Shannon entropy; entropy is independent of how the process is divided into parts. With deterministic collapse, no excess entropy is generated by measuring subsystems, $\delta = 0$.

For the quantum ratchet considered above, the excess entropy is generated from collapse of entangled states when measured locally. A demon that can deterministically predict future collapses from the sequence of preceding outcomes can sufficiently compress the measured outcomes to ensure $\delta = 0$. From Eq. \ref{WnetFinal}, the engine operated by a demon that does not generate excess entropy from collapsing superpositions, will yield $W''_{net} > 0$, in violation of Kelvin's statement (Fig.2).

Quantum effects, such as non-orthogonal states, are on one hand beneficial for operating a heat engine since they can potentially break detailed balance; on the other hand, measuring such states results in generation of excess entropy and added inefficiencies. As we have shown, the second law of thermodynamics precludes the possibility of nonrandom outcomes from measurements of non-orthogonal quantum states. In general, any system with quantum correlations (even without entanglement) has the potential to break detailed balance. However, quantifying the benefits may be more difficult than the above construct. Besides an exercise in connecting fundamental laws, the result that thermodynamic work deficit is not strictly equal to discord might prove useful in identifying the important ingredients for quantum information processing. A quantum ratchet based on decoherence is also of practical interest when generation of excess entropy is tolerable.

\begin{acknowledgments}
The author is grateful to Luca Peliti and Boris Shraiman for helpful discussions. This research was supported in part by the National Science Foundation under Grant No. NSF PHY11-25915.
\end{acknowledgments}

\appendix
\section{Supplementary Information}

We present a detailed analysis of the quantum ratchet engine proposed in the main text of the paper.

\subsection{The subsystem}

The working substance of the engine is comprised of $N \gg 1$ subsystems, each composed of two interacting atoms (A and B) in a box of size $L$. Each atom is a two-level system with energy gap $\hbar\omega$. We set $\hbar=1$ for the remainder of the discussion. The internal state of the atom is represented as a spin-$1/2$ system. The two atoms interact (XY-interaction) with a coupling constant that is dependent on the separation distance of the two atoms.

The Hamiltonian for each subsystem is given by,
\begin{equation}
H = \omega S^z_A + \omega S^z_B + \lambda(\r)(S^+_AS^-_B + S^-_AS^+_B), \label{atomH}
\end{equation}
where $S^z_j = \sigma^z_j/2$ and $S^{\pm}_j = S^x_j \pm i S^y_j$, with $\sigma^{x,y,z}_j$ the Pauli matrix acting on atom $j=A,B$.

The coupling is only turned on when the separation distance of the atoms is below a cutoff.

\begin{equation}
\lambda(\r) = \begin{cases} \lambda, & \mbox{if } |\r| \leq r_0 \\ 0, & \mbox{if } |\r|>r_0 \end{cases}
\end{equation}

Note that there is no kinetic energy term in the Hamiltonian, which implies that the `gas' of two atoms has no pressure. Therefore, there is no energetic cost (gain) in compressing (expanding) the gas by changing the dimensions of the box.

\subsection{Step 1: Initializing the engine}

Initially, the box size is smaller than the interaction range of the atoms, $L_0 < r_0$. The Hamiltonian then has no $\r$ dependence and the coupling is always on. 

\begin{equation}
H_i = \omega S^z_A + \omega S^z_B + \lambda  (S^+_AS^-_B + S^-_AS^+_B), \label{atomHi}
\end{equation}

The atoms are allowed to interact with a thermal radiation field at temperature $T$ to reach equilibrium. Each subsystem is described by density matrix,
\begin{equation}
\rho_{int} = e^{-\beta H_i}/Tr\{e^{-\beta H_i}\}, \label{internalRho}
\end{equation}
 where $\beta = (k_B T)^{-1}$ is the inverse temperature. From here on, we choose our units such that $k_B=1$. The above density matrix corresponds to the internal degrees of freedom of the atomic pair. Next, we assume that the internal degrees of freedom are held fixed (i.e. $\rho_{int}$ does not change) but the system is allowed to equilibrate via the translational degrees of freedom of the atoms.

\subsection{Step 2: Thermalizing translational degrees of freedom}

In this step we assume that with the thermal radiation removed the internal degrees of freedom of the atoms are held fixed. This means that the occupation probability of the energy levels corresponding to Hamiltonian (Eq.\ref{atomHi}) or equivalently the eigenvalues of $\rho_{int}$ are constant. The spacing between the energy levels, however, can change as the Hamiltonian changes as a function of inter-atomic separation distance.

The box size is now expanded to $L \approx 2^{1/3} r_0$. As noted, no energy is extracted from this expansion. We assume that for one atom held fixed, the volume where the coupling in non-zero, $|\r| \leq r_0$, is equal to the volume where coupling is zero, $|\r| > r_0$, so that there is no entropic preference for coupling or uncoupling; there is however an energetic preference. The probability of finding the atomic pair coupled ($|\r| \leq r_0$) is given by,
\begin{equation}
p_i = \bigg\langle \frac{e^{-\beta H_i}}{e^{-\beta H_i}+e^{-\beta H_0}} \bigg\rangle_{\rho_{int}},
\end{equation}
where $H_i$ is the interacting Hamiltonian given above, and $H_0$ is the non-interacting Hamiltonian ($H_i$ with $\lambda = 0$). The expectation value is taken over the internal state density matrix, Eq.\ref{internalRho}. Equivalently, the probability of finding the atomic pair in an uncoupled state is,
\begin{equation}
p_o = \bigg\langle \frac{e^{-\beta H_0}}{e^{-\beta H_i}+e^{-\beta H_0}} \bigg\rangle_{\rho_{int}}. \label{probo}
\end{equation}

The average energy of the uncoupled state is higher than the coupled state. It is therefore less likely to find the atoms with separation distance larger than the interaction cutoff, $|\r| > r_0$. Thermal fluctuations, i.e. from collision of atoms with the walls of the box, however, result in transient excitations, and a non-zero probability of observing the uncoupled state at equilibrium. We can think of each box of two atoms as a two level subsystem, with the excited state corresponding to uncoupled atoms, and the ground state corresponding to coupled atoms. A demon can extract useful work from this system by measuring the state of atoms and extracting the excess energy of the coupled state. However, the demon's observation comes at the cost of increasing the entropy of the demon's memory by an amount equal to the Shannon entropy of $p_i$ and $p_o$. The energy cost of erasing the memory for a cyclic operation exceeds the energy gain from coupling of uncoupled atoms.


Next, we will show that it is possible to extract useful work from thermally excited states of the atomic pairs without measuring the state of each subsystem (box) individually. This is done by collapsing the entangled states of the two atoms by a collective measurement of all the subsystems. Of course, as we will show, the second law is not violated since the collapse of the non-orthogonal states generates excess entropy.

\subsection{Step 3: Collective measurement of the atoms}

The Hamiltonian in Eq.\ref{atomHi} has the following eigenstates: $|g\rangle_A \otimes |g\rangle_B = |gg\rangle$, $|-\rangle = 1/\sqrt{2}(|eg\rangle - |ge\rangle)$, $|+\rangle = 1/\sqrt{2}(|eg\rangle + |ge\rangle)$, and $|ee\rangle$; $|g\rangle$ and $|e\rangle$ correspond to the ground state and excited state of an individual atom, $S^z = \pm 1/2$. The corresponding eigenvalues are respectively $-\omega$, $-\lambda$, $\lambda$, and $\omega$. With the coupling turned off, $\lambda =0$, the energy eigenstates are separable: $|gg\rangle$, $|ge\rangle$, $|eg\rangle$, and $|ee\rangle$, with energy eigenvalues $-\omega$, $0$, $0$, $\omega$ respectively.

The probability of finding internal state $|+\rangle$ in the coupled (interacting) state is given by $p^+_i = 1/(1+e^{\beta \lambda})$. Similarly, the probability of finding internal state $|-\rangle$ in the coupled state is given by $p^-_i = 1/(1+e^{-\beta \lambda})$. The probability of finding these states in the uncoupled state is respectively, $p^+_o = 1- p^+_i$ and $p^-_o = 1- p^-_i$. $|-\rangle$ has a lower energy in the coupled state, and is therefore more likely to be found in this state at equilibrium. $|+\rangle$, however, energetically prefers the uncoupled state and has a higher equilibrium probability of occurring in such state.

One way to decouple the pair of atoms is to measure the spin of one the atoms, for instance atom $A$. Such measurement will collapse the entangled states $|\pm\rangle$ to one of the separable state $|eg\rangle$ or $|ge\rangle$. The act of measurement and collapse of state $|-\rangle$ will cost energy $\lambda$; alternatively, collapse of $|+\rangle$ results in a gain of the same amount of energy.

We first present a simple model of collective spin measurement to convey the conceptual picture, later a more realistic physical implementation using Faraday rotation interaction of the atoms with linearly polarized light is presented.

\emph{Simple model--} A collective quantum non-demolition measurement of the atomic internal states (spins) will determine the total number of $A$ atoms in the excited state; however, it will be impossible to tell which particular boxes contained the excited $A$ atoms. Note that the measurement is `non-demolition' in the sense that the Hamiltonian of probe-interaction commutes with the single-particle Hamiltonian ($S_A^z$) (see section below). Repeated collective measurements will produce the same outcome of number of excited atoms, for uncoupled pairs, or sufficiently short interval between the measurements. Equivalently, the state of the atoms ($|g\rangle$ or $|e\rangle$) is unaltered by the measurement, such that the excited atoms can be used in the next step to extract work.

Let's demonstrate this first for the simpler case when the initial state is a pure state of all subsystems in state $|+\rangle$. The collective measurement will collapse all the entangled states in the subsystems; the resulting state is a superposition of all possible permutations of the excited atoms over all boxes.
\begin{widetext}
\begin{equation}
|0\rangle_p \otimes |+\rangle_1 \otimes \hdots \otimes |+\rangle_N \rightarrow \sum_m \sqrt{p_m} |m\rangle_p \otimes \big( \sum_{i_1,i_2,\hdots,i_m}^N \frac{1}{\sqrt{N \choose m}} |g_1^Ae^B_1,\hdots,e_{i_1}^Ag^B_{i_1},\hdots,e_{i_m}^Ag^B_{i_m},\hdots,g_N^Ae^B_N\rangle \big),
\end{equation}
\end{widetext}
where $p_m$ is the probability of observing $m$ excited atoms. The subscript $p$ denotes the probe state and $1 \hdots N$ the index of the subsystems. The second summation on the right hand side is over all permutations of assigning $m$ excited $A$-atoms to $N$ boxes. After the measurement all the subsystem are in eigenstates of $H_0$.


For the more complicated scenario, where the initial state of each subsystem is the mixed state $\rho_{int}$, the post-measurement state of the system is also a mixed state. However, as in the simpler case, all entangled states in the subsystems will collapse by the measurement of the collective spin. A subsystem can not remain in a superposition of the atomic pair in excited/ground states when the total number of excited $A$-atoms is known (entangled with the state of the probe). Tracing out the $B$-atom states results in a state after the measurement that is a statistical ensemble of all possible permutations of the excited atoms over all the boxes,
\begin{equation}
|0\rangle \langle 0|_p \otimes \rho^A_1 \otimes \hdots \otimes \rho^A_N \rightarrow \sum_m p_m |m\rangle \langle m|_p \otimes \sum_{i_1,i_2,\hdots,i_m}^N  \rho_{i_1,\hdots,i_m},
\end{equation}
where where $\rho^A_i = Tr^B\{\rho_{int}\}$ is the density matrix of atom $A$ in box $i$. The summation in the second term on the right hand site is over all permutations of assigning $m$ excitations to $N$ boxes; each index runs from $1$ to $N$ and no two indices can take on the same value.
\begin{widetext}
\begin{equation}
\rho_{i_1,\hdots,i_m} = \frac{1}{{N \choose m}} |g_1,\hdots,e_{i_1},\hdots,e_{i_m},\hdots,g_N\rangle \langle g_1,\hdots,e_{i_1},\hdots,e_{i_m},\hdots,g_N|_A.
\end{equation}
\end{widetext}
$\rho^A_i$ is always diagonal. Denote the probability of finding atom $A$ in the excited state as $p^A_e$ and in the ground state as $p^A_g=1-p^A_e$.The probe before the measurement is in state $|0\rangle_p$. After interacting with the system, the probe is in state $|m\rangle_p$ (where $m$ denotes the number of excitations) with probability $p_m$. The number of excited atoms is not affected by the interaction of the probe since it is a non-demolition measurement. A measurement of the probes state with an outcome $m'$ implies that the $A$-atoms are described by density matrix,
\begin{widetext}
\begin{equation}
 \sum_{i_1,i_2,\hdots,i_{m'}}^N  \frac{1}{{N \choose m'}} |g_1,\hdots,e_{i_1},\hdots,e_{i_{m'}},\hdots,g_N\rangle \langle g_1,\hdots,e_{i_1},\hdots,e_{i_{m'}},\hdots,g_N|_A.
\end{equation}
\end{widetext}
The atomic pairs in each box are no longer entangled in the post-measurement state above. Having observed the state the $A$ atoms has collapsed the entangled states  $|\pm\rangle$; the resulting delocalized excitations imply that now all the boxes are entangled; this entanglement, however, does not affect the energy of the system. There is an energy cost to the collapse of $|\pm\rangle$ for the atomic pairs with non-zero coupling. On average, the work required for conducting this measurement is given by,
\begin{equation}
W_p = N \lambda ( p_- p^-_i  - p_+ p^+_i), 
\end{equation} 
where $p_\pm$ is the equilibrium probability of observing the entangled states $|\pm\rangle$ (Eq.\ref{internalRho}). $p^{\pm}_i$ is the probability of finding state $|\pm\rangle$ in the coupled state, as defined above. $W_p$ is not a work cost in the cycle. The energy put in the system in collapsing the energetically favorable entangled states will be retrieved in the next step, when work is extracted from the atoms.

The actual cost of the measurement is in the information gained, which incurs an erasure penalty. It is easy to show that the information penalty from the probe's measurement is negligible. First, since we have knowledge of the initial density matrix of the atoms $\rho_{int}$ (having prepared it in the first step), we can easily calculate the average number of excited $A$ atoms, $Np^A_e$. The only information gained from the measurement is fluctuations around this average due to a finite $N$. Number of excitations is given by a binomial distribution with variance $\sigma^2 = Np^A_ep^A_g$. The information gained per box from the measurement is the entropy of the binomial distribution,
\begin{equation}
H_p = \frac{1}{2N} \log_2(2\pi e \sigma) \sim \log(N)/N
\end{equation}
For $N \gg 1$, the information gained from the finite-size fluctuations is negligible. The collective measurement seems to have given us everything for free. The collapse of the entangled states effectively decoupled all the atoms, raising all the subsystems to their `excited' state. The work cost of the measurement --corresponding to raising the coupled ground state pair to uncoupled excited states, is retrieved in the next step when the demon measures the atoms individually. The non-local entanglement between all the boxes due to delocalization of the excitations will also collapse when each atom is measured individually; there is no energetic cost to causing this collapse. Moreover, the entropic cost of the information gained by the collective measurement is negligible for large system sizes. Naively, it might seems that this procedure allows us to extract useful work from thermal fluctuations that resulted in uncoupled atoms in the previous step. However, as we will show, the collapse of the entangled states, encoding the quantum correlations between the two atoms in each box, results in generation of excess entropy. When the demon measures the state of all the individual atoms to extract work, it has to be pay an erasure cost for this excess entropy, which exceeds the gain from `ratcheting' of thermal fluctuations.

\emph{Faraday rotation interaction--} A possible implementation of the quantum non-demolition collective measurement discussed above is using Faraday rotation interactions of the atoms with a linearly polarized light pulse. Define the collective spin operator of $A$ atoms as, $\tilde{S}^z_A = 1/\sqrt{N/2} \sum_{i=1}^N S_{A,i}^z$, where the $i$ summation runs over all boxes. The y and z components of the normalized Stokes operator of a pulse of probe light of duration $t$ is defined as, $\tilde{L}_y = (i\sqrt{2N_L})^{-1}\int_0^t(a^\dagger_+a_--a^\dagger_- a_+)dT$, $\tilde{L}_z = (\sqrt{2N_L})^{-1}\int_0^t(a^\dagger_+a_+-a^\dagger_-a_-)dT$, where $N_L$ is the average number of photons in the pulse, and $a_\pm$ is the annihilation operator of $\sigma_\pm$ circularly polarized light mode \cite{Duan00}. The Hamiltonian for Faraday rotation interaction is given by, $H_{FR} = \alpha \tilde{L}_z\tilde{S}^z_A$, for some real constant $\alpha$ \cite{Takahashi99}. The interaction does not modify the z-component of the collective spin, $[H_{FR},\tilde{S}^z_A] = 0$, and therefore constitutes a quantum non-demolition measurement. The y-component of the Stokes operator evolves under this interaction to $\tilde{L}_y \rightarrow \tilde{L}_y +\kappa\tilde{S}^z_A$, for come constant $\kappa$. A measurement of $\tilde{L}_y$ will project the system into an eigenstate of the collective spin operator $\tilde{S}^z_A$, in analogy to the fixed excitation-number $|m'\rangle$ state of the probe discussed above. For an example of experimental implementation of this scheme see \cite{Kuzmich00}.

\subsection{Step 4: Demon measures the subsystems}

In this step, the demon measures the internal state (spin) of every atom and extracts useful work. Following the collective measurement, the size of each box is reduced to that of Step $1$, $L < r_0$, ensuring that two atoms are always interacting. With no entangled pairs, there is no energetic cost for reducing the size of the subsystems. This ensures that the demon's measurement of the atoms contains no information on whether the atoms were in a coupled or uncoupled state.

The demon measures the state of each atom in every subsystem. Let's assume without loss of generality that atom $A$ is measured first in box 1. As discussed in the main text, the demon gains information $S(\rho^A_1) = -Tr \{\rho^A_1 \ln \rho^A_1 \}$. The demon then measures the state of atom $B$. Because of correlations, information gained from this measurement is not independent of the measurement outcome of atom A. Entropy of measurement of $B$ is given by $S(\rho^B_1|\Pi_A) = \Sigma_i p_{\pi_A = a_i} S(\rho^{B|\pi_{a_i}}_1)$, where $\Pi_A$ is a complete projective measurement on A; $p_{\pi_A = a_i} = Tr\{\rho^A_1 \Pi_{A=a_i}\}$ is the probability of outcome $a_i$ from measuring subsystem A; and $\rho^{B|\pi_{a_i}}_1 = Tr^A\{\rho_1  \Pi_{A=a_i} \}/p_{\pi_A = a_i}$ is the state of B after the measurement on A. As noted in the main text, the total entropy of the demon's measurement is greater than the entropy of the initial state $\rho_{int}$ due to the random collapse of the entangled states by the collective measurement. The excess entropy is given by,
\begin{equation}
\delta = S(\rho^A_1) + S(\rho^B_1|\Pi_A) - S(\rho_{int}). \label{discord}
\end{equation}
This is the entropic price of having collapsed the superpositions in each subsystem.

From the single-atom density matrices, individual atoms appear to be at an effectively higher temperature,
\begin{equation}
T_{eff} = -\omega/\ln (1/\rho^{A,B}_{2,2} -1),
\end{equation}
where $\rho^{A,B} = Tr_{B,A}\{\rho_{int}\}$, and the subscript $2,2$ refers to the corresponding element ($|e\rangle\langle e|$) of the single-atom density matrix. For the above system $T_{eff} > T$. The atoms are effectively hotter after the collapse of the entangled states. The effective higher temperature is due to the work put in the system during the collective measurement that uncoupled the atomic pairs.

\subsection{Net work extracted}

The demon extracts work from each box by using the following procedure: 1) Demon measures the state of each atomic pair, reducing the subsystem's entropy to zero. 2) The subsystem is placed in contact with the thermal reservoir and allowed to isothermally and reversibly expand to its initial state, which is characterized by $\rho_{int}$. Work is extracted during this quasi-static reversible expansion. 3) The demon erases the information obtained from measuring each subsystem.

The net work extracted from the collective measurement and the actions of the demon can be calculated using the first law of thermodynamics.
\begin{equation}
\Delta U = W + Q, \label{FirstLaw}
\end{equation}
where $\Delta U$ is the change in internal energy, $W$ the net work performed by the system, and $Q$ the net heat flow, per box.

Let's analyze each term of the above equation separately. The change in the internally energy is zero for the atoms-pairs that where in the coupled state (ground state of the effective two-level system of each atomic pair) at the time of the collective measurement, since the initial and final states are the same. However, the story is different for the thermally excited states. Thermally uncoupled $|-\rangle$ states are energetically beneficial, since they require no work expenditure during the collective measurement, but result in a reduction of the internal energy by $\lambda$ after the reversible expansion to the initial coupled state. Similarly, uncoupled $|+\rangle$ states require no work during the collective measurement. However, they increase the internal energy by $\lambda$ when restored to the initial coupled state.

The average net change in the internal energy of a subsystem in a cycle is given by
\begin{equation}
\Delta U = \lambda(p_- p^-_o - p_+ p^+_o), 
\end{equation}
where $p^\pm_o$ is the probability of finding internal state $|\pm \rangle$ in the uncoupled state (see above). $p_{\pm}$ is the equilibrium probability of internal state $|\pm \rangle$ deduced from density matrix $\rho_{int}$ (Eq.\ref{internalRho}). This non-zero change in internal energy corresponds exactly to the energy `ratcheted' from thermal kicks in the system. In the limit of very small coupling, $\lambda \to 0$, $\Delta U \sim \lambda^2$. The second law of thermodynamics precludes conversion of this energy into useful work by completing a cycle. Below, we show that the entropy generation from measurement-induced collapse of the non-orthogonal states compensates the work potential of the ratchet.

The last term in Eq.\ref{FirstLaw} is the net heat flow out of the system (to be consistent with our sign convention) per subsystem. Heat flow into the system during the isothermal expansion following the demon's measurement is given by,
\begin{equation}
Q_{in} \leq T S(\rho_{int}),
\end{equation}
where $S$ denotes the von Neumann entropy. The inequality becomes an equality when the isothermal expansion occurs reversibly and quasi-statically. Any irreversibility in the expansion implies entropy generation without the corresponding heat flow, which in turn implies a lower extractable work. 

The heat flow out of the system is set by the erasure work.
\begin{equation}
Q_{out} = TS(\rho_{int}) +T\delta,
\end{equation}
where $\delta$ (Eq.\ref{discord}) is the excess entropy generated from the probabilistic collapse of the entangled states induced by the collective measurement.

Putting everything together, the net work performed by the system must satisfy
\begin{equation}
W \leq \lambda(p_- p^-_o - p_+ p^+_o) + T S(\rho_{int}) - TS(\rho_{int}) - T\delta.
\end{equation}

With the assumption that the isothermal expansion is reversible,
\begin{equation}
W_{cyc} = \lambda(p_- p^-_o - p_+ p^+_o) - T\delta. \label{Wcyc}
\end{equation}

Without loss of generality, the energy units are selected such that $\omega =1$. Subsequently, $W_{cyc} \leq 0$, for all values of $\lambda$ and $T$; in accordance with the second law of thermodynamics, ensuring that no net work is extracted from the cycle. As discussed in the main text, if the outcomes of the collapse of the entangled states were nonrandom, then $\delta = 0$, allowing the demon to extract work from each cycle, in direct violation of Kelvin's statement.



\begin{thebibliography}{10}

\bibitem{Kelvin} Thomson, W. (Lord Kelvin) An account of Carnot's theory of the motive power of heat; with numerical results deduced from Regnault's experiments on steam. {\it Trans. Roy. Soc. Edinburgh} {\bf 16}, 541-574 (1849).

\bibitem{Bohr} Bohr, N. The quantum postulate and the recent development of atomic theory. {\it Nature} {\bf 121}, 580Ð590 (1928).

\bibitem{Born} Born, M., Zur Quantenmechanik der StossvorgŠnge. {\it Zeits. Phys.} {\bf 37}, 863-867 (1926).

\bibitem{Zeilinger00} Jennewein, T., Achleitner, U., Weihs, G., Weinfurter, H. \& Zeilinger, A. A fast and compact quantum random number generator. {\it Rev. Sci. Instrum.} {\bf 71}, 1675 (2000).

\bibitem{ExpTestRand1} Erber, T. \& Putterman, S. Randomness in quantum mechanicsÑnature's ultimate cryptogram? {\it Nature} {\bf 318}, 41 (1985).

\bibitem{ExpTestRand2} Silverman, M. P., Strange, W., Silverman, C. \& Lipscombe, T. C. Tests for randomness of spontaneous quantum decay. {\it Phys. Rev. A} {\bf 61}, 042106 (2000).

\bibitem{ExpTestRand3} Berkeland, D. J., Raymondson, D. A. \& Tassin, V. M. Tests for nonrandomness in quantum jumps {\it Phys. Rev. A} {\bf 69}, 052103 (2004).

\bibitem{Calude08} Calude, C. S. \& Svozil, K. Quantum randomness and value indefiniteness. {\it Adv Sci Lett} {\bf 1}, 165-168 (2008). 

\bibitem{Zurek09a} Zurek, W. H. Quantum darwinism. {\it Nature Phys.} {\bf 5}, 181 (2009)

\bibitem{Zurek09b} Zurek, W. H. Entanglement Symmetry, Amplitudes, and Probabilities: Inverting Born's Rule. {\it Phys. Rev. Lett.} {\bf 106}, 250402 (2011).

\bibitem{ZurekSzilard} Zurek, W. H. Maxwell's Demon, Szilard's Engine and Quantum Measurements. e-print arXiv:quant-ph/0301076v1.

\bibitem{Nori07} Quan, H. T. , Liu, Y.-X., Sun,  C. P. \& Nori, F. Quantum thermodynamic cycles and quantum heat engines. {\it Phys. Rev. E} {\bf 76}, 031105 (2007).

\bibitem{Kieu04} Kieu, T. D. The second law, Maxwell's demon, and work derivable from quantum heat engines. {\it Phys. Rev. Lett.} {\bf 93}, 140403 (2004).

\bibitem{Ueda11} Kim, S.W, Sagawa, T., De Liberato, S. \& Ueda, M. Quantum Szilard Engine. {\it Phys. Rev.Lett.} {\bf 106}, 070401 (2011).

\bibitem{Oppenheim02} Oppenheim, J., Horodecki, M., Horodecki, P. \& Horodecki, R. Thermodynamical approach to quantifying quantum correlations. {\it Phys. Rev. Lett.} {\bf 89}, 180402 (2002).

\bibitem{Zurek03} Zurek, W. H. Quantum discord and MaxwellÕs demons. {\it Phys. Rev. A} {\bf 67}, 012320 (2003).

\bibitem{Feynman} Feynman, R. P., Leighton, R. B. \& Sands, M. {\it The Feynman Lectures on Physics , Vol. I, Chap. 46.} (Addison-Wesley, Reading, MA, 1966).

\bibitem{Maxwell} Maxwell, J. C. {\it Theory of Heat, 4th ed., p. 328-329} (Longmans, Green \& Co., London, 1875).

\bibitem{Szilard} Szilard, L. †ber die Entropieverminderung in einem thermodynamischen System bei Eingriffen intelligenter Wesen. {\it Z. Phys.} {\bf 53}, 840-856 (1929).

\bibitem{LBa} Landauer, R. Irreversibility and heat generation in the computing process. {\it IBM J. Res. Dev.} {\bf 3}, 183 (1961)

\bibitem{LBb} Bennett, C. H. The thermodynamics of computation --a review. {\it Int. J. Theor. Phys.} {\bf} 21, 905 (1982).


\bibitem{vonNeumann55} J. von Neumann, {\it Mathematical Foundations of Quantum Mechanics} (Princeton University Press, Princeton, Eng. translation by R. T. Beyer, 1955).

\bibitem{Lloyd97} Lloyd, S. Quantum-mechanical Maxwell's demon. {\it Phys. Rev. A} {\bf 56}, 3374 (1997).

\bibitem{Scully03} Scully, M. O., Zubairy, M. S., Agarwal, G. S. \& Walther, H. Extracting work from a single heat bath via vanishing quantum coherence. {\it Science} {\bf 299}, 862 (2003).


\bibitem{Ueda08} Sagawa, T. \& Ueda, M. Second Law of Thermodynamics with Discrete Quantum Feedback Control. {\it Phys. Rev. Lett.} {\bf 100}, 080403 (2008).

\bibitem{Vitelli01} Plenio, M. B. \& Vitelli, V. The physics of forgetting: Landauer's erasure principle and information theory. {\it Contemporary Physics} {\bf 42}, 25 (2001).

\bibitem{Vedral10b} Vedral, V. The elusive source of quantum speedup. {\it Found. Phys.} {\bf 40}, 1141 (2010).

\bibitem{Zurek01a} Zurek, W. H. {\it Ann. Phys.} (Leipzig) {\bf 9}, 855 (2000)

\bibitem{Zurek01b} Ollivier, H. \& Zurek, W. H. Quantum discord: A measure of the quantumness of correlations. {\it Phys. Rev. Lett.} {\bf 88}, 017901 (2001).

\bibitem{Vedral01} Henderson, L. \& Vedral, V. Classical, quantum and total correlations. {\it Journal of Physics A} {\bf 34}, 6899 (2001).

\bibitem{Vedral10} Modi, K., Paterek, T., Son, W., Vedral, V. \& Williamson, M. Unified view of quantum and classical correlations. {\it Phys. Rev. Lett.} {\bf 104}, 080501 (2010).



\bibitem{Smoluchowski} von Smoluchowski, M. Experimentell nachweisbare, der Ÿblichen Thermodynamik widersprechende MolekularphŠnomene. {\it Phys. Z.} {\bf XIII}, 1069 (1912).

\bibitem{Magnasco} Magnasco, M. Forced thermal ratchets. {\it Phys. Rev. Lett.} {\bf 71}, 1477 (1993).

\bibitem{Astumian} Astumian, R. D. Thermodynamics and kinetics of a Brownian motor. {\it Science} {\bf 276}, 917 (1997).

\bibitem{Lukin} Lukin, M. D. Colloquium: Trapping and manipulating photon states in atomic ensembles. {\it Rev. Mod. Phys.} {\bf 75}, 457-472 (2003).

\bibitem{Kolmogorova} Solomonoff, R. J. A formal theory of inductive inference. Part I. {\it Inf. Control} {\bf 7}, 1 (1964).

\bibitem{Kolmogorovb} Kolmogorov, A. N.  Three approaches to the quantitative definition of information. {\it Inf. Transmission} {\bf 1}, 3 (1965). 

\bibitem{Kolmogorovc} G. J. Chaitin, {\it Algorithmic Information Theory} (Cambridge University Press, Cambridge, England, 1987).

\bibitem{Zurek89} Zurek, W. H. Algorithmic randomness and physical entropy. {\it Phys. Rev. A} {\bf 40}, 4731 (1989).

\bibitem{Duan00} Duan, L. M.  et al., Phys. Rev. Lett. {\bf 85}, 5643 (2000).

\bibitem{Takahashi99} Takahashi, Y. et al., Phys. Rev. A {\bf 60}, 4974 (1999).

\bibitem{Kuzmich00} Kuzmich, A., Mandel, L. \& Bigelow, N. P. Phys. Rev. Lett. {\bf 85}, 1594 (2000).


\end{thebibliography}
\end{document}